\documentclass[12pt]{article} 
\usepackage{tabularx}
\usepackage[latin1]{inputenc}
\usepackage{pstricks,pst-node,pst-text,pst-3d}
\usepackage{amsmath,amsfonts,amssymb}
\usepackage{verbatim}
\usepackage{subfigure}
\usepackage{amssymb,fontenc,times,mathptmx,color,graphicx,ifpdf}
\usepackage{epsfig}
\usepackage{makeidx}
\usepackage{multicol}
\font\openface=msbm10 at12pt 
\newcommand{\be}{\begin{equation}}
\newcommand{\en}{\end{equation}}
\newcommand{\bea}{\begin{eqnarray}}
\newcommand{\ena}{\end{eqnarray}}

\newcommand{\ex}{\mathrm{e}}
\newcommand{\ud}{\mathrm{d}}

\renewcommand{\vector}[1]{\overrightarrow{#1}}
\renewcommand{\bar}[1]{\overline{#1}}

\newcommand{\plotone}[1]{\epsfig{file={#1},height=12cm,width=8cm,angle=-90}}
\newcommand{\plottwo}[1]{\epsfig{file={#1},height=10cm,width=6cm,angle=-90}}

\newcommand{\pdd}{{\bf{PDD}}}
\newcommand{\ptf}{{\bf{PTF}}}
\newcommand{\uma}{{\bf{UMA}}}
\newcommand{\mma}{{\bf{MMA}}}
\newcommand{\nn}{\nonumber}
\newcommand{\s}[1]{\mathbb{S}^{#1}}
\newcommand{\sfuzzy}[1]{\mathbb{S}^{#1}_{F}}

\newcommand{\real}[1]{\mathbb{R}^{#1}}
\newcommand{\y}[2]{Y_{_{#1 #2}} \left( \vartheta , \varphi \right)}
\newcommand{\ys}[2]{Y_{_{#1 #2}}^{*} \left( \vartheta , \varphi \right)}
\newcommand{\yp}[2]{\hat{Y}_{_{#1 #2}}}
\newcommand{\an}{\left( \vartheta , \varphi \right)}
\newcommand{\herm}[1]{#1 \times #1}
\newcommand{\matN}[1]{Mat_{_{#1}}}
\newcommand{\su}[1]{SU \left( #1 \right)}

\newcommand{\traza}[1]{\mathrm{Tr} \left[ #1 \right]}

\newcommand{\mbf}{\textbf{MBF}}
\begin{document}
\title{Simulation of a scalar field on a fuzzy sphere}
\author{Fernando Garc\'{\i}a Flores \\ {\it Depto. de F\'\i sica,
    Cinvestav, Apartado Postal 70-543, M\'exico D.F. 0730, MEXICO
}\\Xavier Martin \\
{\it Université François Rabelais, Tours. Laboratoire de Mathématiques et
Physique Théorique} \\ {\it CNRS, UMR 6083. Fédération de Recherche Denis
Poisson (FR 2964)} \\ Denjoe
  O'Connor \\ {\it School of
 Theoretical Physics, DIAS, 10 Burlington Road, Dublin 4, IRELAND}}
\maketitle \abstract{ The $\phi^4$ real scalar field theory on a fuzzy
  sphere is studied numerically. We refine the phase diagram for this
  model where three distinct phases are known to exist: a uniformly
  ordered phase, a disordered phase, and a non-uniform ordered phase
  where the spatial $SO(3)$ symmetry of the round sphere is
  spontaneously broken and which has no classical equivalent. The
  three coexistence lines between these phases, which meet at a triple
  point, are carefully located with particular attention paid to the
  one between the two ordered phases and the triple point itself.  In
  the neighbourhood of the triple point all phase boundaries are well
  approximated by straight lines which, surprisingly, have the same
  scaling.  We argue that unless an additional term is added to
  enhance the effect of the kinetic term the infinite matrix limit of
  this model will not correspond to a real scalar field on the
  commutative sphere or plane.}
\vfill\eject
\section{Introduction}
The fuzzy approximation scheme \cite{fuzzy} consists in approximating
the algebra of functions on a manifold with a finite dimensional
matrix algebra and in principle provides a regularization of field
theory on this space which can be used as an alternative to
discretising the underlying space as is done in lattice field theory.
Both the two--dimensional commutative \cite{Medina:2005su} and Moyal
planes can be viewed as the limits of a fuzzy sphere of infinite
radius.

Here, we study a real scalar field, $\phi$, with $\phi^4$ interaction,
in the fuzzy approach using Monte Carlo simulations. This becomes a
Hermitian matrix model on the fuzzy sphere.  The study reveals that
the model has three distinct phases: (i) A disordered phase; (ii) a
uniformly ordered phase and (iii) a non-uniformly ordered phase
assimilated to a striped phase \cite{HS,Ambjorn,BHN}. We find the
collapsed phase diagram and in particular we calculate the uniform
ordered/non-uniform ordered line that was absent in
\cite{Martin:2004un} and locate the triple point where the three
phases meet. As the mass parameter varies, the non-uniformly ordered
phase is absent for sufficiently small coupling, but as the coupling
is increased this new phase opens up between the disordered and
uniformly ordered phases. The three phases meet at a triple point.

The transition from the disordered to the non-uniformly ordered phase
can also be identified with the one-cut to two-cut transition in
matrix model theory
\cite{O'Connor:2007ea,O'Connor:2007ui,Steinacker:2005wj}.  This
transition line merges with the predicted curve obtained from the
quartic potential of the single trace pure matrix transition for
sufficiently large couplings, i.e. sufficiently above the triple
point.  The qualitative features of the phase diagram are governed by
this triple point. The presence of the non-uniformly ordered phase is
the principal feature that distinguishes the phase diagram of the
fuzzy model from its commutative counterpart.

A preliminary version of these results were presented in Lattice 05
and appeared in \cite{GarciaFlores:2005xc}. The principal aspects of
these results have been confirmed in subsequent studies by Panero
\cite{marco,Panero:2006cs} and Das et al. \cite{Das:2007gm}.

The current study could be relatively easily repeated for a Hermitian
scalar field on other fuzzy spaces. The simplest extension would be to
fuzzy $\s{2}\times\s{2}$ or to fuzzy
$\mbox{\openface CP}^{\rm N}$ \cite{Balachandran:2001dd}.  Fuzzy
versions of $\s{3}$ and $\s{4}$ are also accessible \cite{spheres} and
will hopefully be studied in the near future. In all cases, the
structure of the phase diagram should be similar, although there is no
guarantee that all coexistence lines will collapse with a consistent
scaling as happens for the two dimensional sphere.  One prediction for
the general case is that the disordered non-uniformly ordered line
will always be present for sufficiently large coupling and will again
merge with the pure one-cut two-cut transition for the pure matrix
model.

In section $2$ we review the construction of the fuzzy sphere and in
section $3$ we present the model, section $4$ describes the Metropolis
algorithm, section $5$ studies limiting models such as the lowest
matrix size (two by two) and the pure matrix model, section $6$
describes the observables and simulations, particularly
the specific heat which we use to locate transitions.  Section $7$
gives our main results and describes the collapsed phase diagram and
locates the triple point.  Section $8$ gives our conclusions. The
paper ends with some technical appendix for the  optimization of the
simulations. 

\section{The fuzzy sphere}\label{chap:fuzzysphere}
Before introducing the fuzzy sphere, let us look at some basic
properties of the \emph{ordinary continuum
  $2-$sphere}\index{Sphere!continuum $n-$}.  A 2-sphere centered on
the origin, with radius $R$, embedded in $\real{3}$, denoted simply
$\s{2}$, can be defined as the set of points $ \left( x_{_{1}},
  x_{_{2}}, x_{_{3}} \right)$ in $ \real{ 3 }$ such that $x_{ _{1} }^{
  2 } + x_{ _{2} }^{ 2 } + x_{ _{3} }^{ 2 } = R^{ 2 }$. It can  also
  be expressed by the angles $(\vartheta,\varphi)$ of spherical coordinates.

Taking two elements of the algebra, $f \an$ and $g \an$, we define their
\emph{inner product}\index{Inner product} as
\begin{equation}\label{eq:scalarprod}
\left< f | g \right> = \int_{\s{2}} \ud \Omega\, f^{*} \an g \an, 
\end{equation}
and their \emph{norm} as
\begin{equation} \label{eq:norm} 
  \left\| f  \right\|^2 =  \left< f  | f \right>  = \int_{  \s{2}} \ud
  \Omega\, |f \an |^{2}  .
\end{equation}
where ${\displaystyle \int_{\s{2}} \ud \Omega = \int_{0}^{2 \pi} \ud \varphi
\int_{ 0 }^{  \pi } \ud \vartheta \sin  \left( \vartheta \right)}$. The
norm must be finite for  any element of the algebra (square integrable
functions).     Both     equations,     (\ref{eq:scalarprod})     and
(\ref{eq:norm}), define the \emph{Hilbert space}\index{Hilbert space}
$\mathcal{H}$ which allows us to quantize the theory.

In  general, the \emph{Laplace  operator}\index{Laplacian} contains
information on the geometry of the space, \emph{i.e.}
it depends on the metric as $\nabla^2 \cdot = \frac{1}{ \sqrt{ |g| } }
\partial_{i}  \sqrt{  |g| }  \partial^{i}  \cdot$,  where  $g$ is  the
determinant  of  the metric  tensor  $g_{\mu  \nu}$  on Riemannian  and
pseudo-Riemannian  manifolds~\cite{laplacian}.   In particular,  the  Laplacian on  the
sphere is  $\nabla^2 =  \frac{1}{\left( R \sin \vartheta  \right)^{2} }
\frac{  \partial^2}{  \partial  \varphi^2}  + \frac{1  }{  R^{2}  \sin
  \vartheta}  \frac{  \partial}{  \partial  \vartheta  }  \left(  \sin
\vartheta  \frac{  \partial }{  \partial  \vartheta  } \right)$.   The
eigenfunctions  of  this  operator   are  the  spherical  harmonics  $
\y{\ell}{m}$ with $ \ell = 0,1,2,\dots$ and $m=-\ell , -\left( \ell -1
\right) ,  \dots ,  \left( \ell -1  \right) ,  \ell $ which  come as
solutions of the  Helmoltz equation $\Delta f +l(l+1)f = 0$ on the
unit sphere. 

A convenient basis to describe any function on the sphere is given
by  these \emph{spherical  harmonics}\index{Spherical harmonics}  $ \y{
  \ell }{ m } $ since they form a complete set of orthonormal functions and
thus,  any  square-integrable function  can  be  expanded  as a  linear
combination of these.
\begin{equation}\label{eq:expansionf}
f \an =  \sum_{\ell=0}^{ \infty }\ \sum_{m = -\ell }^{  + \ell } c_{_{\ell
    m}} \y{\ell}{m}, 
\end{equation}
where the orthonormalization condition
\begin{equation}\label{eq:orthonormalization}
\int_{\s{2}}  \ud \Omega\, \ys{\ell}{m}  \y{\ell^{\prime}}{m^{\prime}} =
\delta_{\ell \ell^{\prime}}\delta_{m m^{\prime}}, 
\end{equation}
allows us to compute the $c_{_{\ell m}}$ coefficients as
\begin{equation}\label{eq:coefficiets}
  c_{_{\ell m}} = \int_{\s{2}} \ud \Omega\, \ys{\ell}{m} f \an. 
\end{equation}

We are now ready to define the \emph{fuzzy sphere}\index{Fuzzy
  sphere}\index{Sphere!fuzzy} $\sfuzzy{2}$ of radius $R$
\cite{Landi:1997sh,Douglas:2001ba,Madore:1991bw}. It is a
non-commutative space defined in terms of the $N \times N$ matrix
operators $\left( \hat{ x }_{ 1 }, \hat{ x }_{ 2 },\hat{ x }_{ 3 }
\right)$ subject to the relations
\begin{equation}\label{eq:noncommutative} 
  \hat{x}^{2}_{1} + \hat{x}^{2}_{2}+\hat{x}^{2}_{3}=R^{2} \hat{
  1}, \quad \text{and} \quad \left[ \hat{x}_{i} , \hat{x}_{j} \right] = i
  \epsilon_{ijk}  \frac{  2 R  }{  \sqrt{  N^{2}  - 1  }}
  \hat{x}_{k}=i\epsilon_{ijk} \Theta \frac{\hat{x}_k}{R},
\end{equation}
with $\Theta=2R^2/\sqrt{N^2-1}$ and $\epsilon_{ijk}$ the totally
antisymmetric unit tensor.
The  operators $\hat{x}_{i}$ can  be related  to the  angular momentum
operators  $\hat{L}_{i}$   in  their  irreducible   representation  of
$\su{2}$ of size $ \left( 2 \ell +1 \right)$ with the formula 
\begin{equation}\label{eq:relation}
\hat{x}_{i} =\frac{ 2 R }{ \sqrt{ N^{2} - 1 }} \hat{L}_{i}=\frac{\Theta}{R} \hat{L}_i, 
\end{equation}
where the relation between the  matrix size $N$ and the representation
 of  the  angular  momentum  $\ell$  is   given  by  $  N=  2  \ell  +
 1$.     Replacing     the     equation     (\ref{eq:relation})     in
 (\ref{eq:noncommutative}) we recover the angular momentum algebra.

In the  table \ref{tab:fuzzyspaces} we  show some limits of  the fuzzy
sphere in  terms of the matrix size  $N$ and the radius  of the sphere
$R$.  In that  way, the  fuzzy sphere  contains some  other  spaces as
limits of the matrix size and its radius.

\begin{table}[tbp]
  \begin{center}
    \begin{tabular}{||c|c|c|l||}
      \hline
      \hspace{0.5cm}$N$\hspace{0.5cm} & \hspace{0.5cm}$R$\hspace{0.5cm}
      &  $\Theta$ & \hspace{0.5cm}Limit
      \hspace{0.5cm} \\
      \hline
      $N_0$  & $R_0$  &  $\frac{  2 R^2  }{  \sqrt{ N^{2}  -  1 }}$  &
      Fuzzy sphere ($\sfuzzy{2}$)\\ 
      $+\infty$ & $R_0$ & $0$ & Commutative sphere ($\s{2}$)\\
      $+\infty$ & $+\infty$ & $0$ & Commutative plane ($\real{2}$)\\
      $+\infty$ & $+\infty$ & $\Theta_0$ & Moyal plane \\
      \hline
      
    \end{tabular}
  \end{center}\caption{Some   spaces   as    limits   of   the   fuzzy
      sphere.}\label{tab:fuzzyspaces} 
\end{table} 

From the algebra of matrices of size $N$, denoted $\matN{N}$,
generated by the position operators $\hat{x}_i$ in
(\ref{eq:noncommutative}), one can define a \emph{Hilbert
space}\index{Hilbert space}, by introducing an \emph{inner
product}\index{Inner product}.  To do that, consider two elements
of the algebra $\matN{N}$ denoted $\phi$ and $\psi$, their scalar
product and associated \emph{norm}\index{Norm} are defined by
\begin{equation}
  \langle   \phi   |  \psi   \rangle   =  \frac{   4   \pi   }{  N   }
  \traza{\phi^{\dagger } \psi} ,\ 
  \| \phi \|^2  = \langle  \phi |  \phi \rangle =  \frac{ 4  \pi }{  N }
     \traza{\phi^{ \dagger } \phi } , 
\end{equation}
where the normalization was chosen so that the unit matrix $\hat{1}$
and the constant function $1$ on the sphere have the same norm.

The geometry of the spaces is given through derivation operators. In
the case of $\sfuzzy{2}$, the derivations $\mathcal{L}_i$ correspond
to the adjoint action $ [ \hat{L}_i,\cdot ] $ of the angular momentum
operators $\hat{L}_{i}$ of $\su{2}$. The Laplacian is then deduced as
\begin{equation}
\mathcal{L}^2 \phi= \mathcal{L}_i \mathcal{L}_i \phi= [ \hat{L}_i, [
\hat{L}_i ,\phi ] ] .
\end{equation}

Similar to the expansion (\ref{eq:expansionf}) of a function $f \an$
on $\s{2}$, a convenient basis to expand any $\herm{N}$ matrix $\phi$
on $\sfuzzy{2}$ is the \emph{polarization tensor}
basis\index{Polarization tensor}.  The polarization tensors are
denoted by $\yp{\ell}{m}$ with $0 \le \ell \le \left( N - 1 \right)$
and $ - \ell \le m \le + \ell $, and are defined as the simultaneous
eigenvectors of the laplacian $\mathcal{L}^2$ and axial angular
momentum $\mathcal{L}_3$:
\begin{equation}
\mathcal{L}^2 \hat{Y}_{lm}=l(l+1)\hat{Y}_{lm},\ \mathcal{L}_3
\hat{Y}_{lm}=m\hat{Y}_{lm},
\end{equation}
and we see that $\mathcal{L}^2$ is a cut--off version of $-\nabla^2$.
They are normalised to form an orthonormal basis of $\matN{N}$
\begin{equation}\label{eq:orthopolarization}
  \frac{4  \pi  }{ N } Tr  \left[  \yp{\ell}{m}^{\dagger}
    \yp{\ell^{\prime}}{m^{\prime}}     \right]     =    \delta_{_{\ell
    \ell^{\prime}}} \delta_{_{m m^{\prime}}}, 
\end{equation}
and transform simply under complex conjugation
\begin{equation}\label{eq:conjupolarization}
\yp{\ell}{m}^{\dagger} = \left( -1 \right)^{m}  \yp{\ell}{\ -m}.
\end{equation}

The expansion of $\phi$ in $\yp{\ell}{m}$ is given by
\begin{equation}\label{eq:fuzzyexpansion}
  \phi   =  \sum_{\ell=0}^{N-1}\ \sum_{m=-\ell}^{+\ell}   c_{_{\ell  m}}
  \yp{\ell}{m} ,
\end{equation}
where the coefficients can be  computed by means of the orthonormality
condition 
\begin{equation}\label{eq:coefis}
 c_{lm} = \frac{4 \pi }{ N } \traza{ \yp{\ell}{m}^{\dagger} \phi }.  
\end{equation}

 \section{Real scalar field on a fuzzy sphere}\label{sec:modelonafuzzysphere}
Before introducing the real scalar field theory on the fuzzy sphere,
let us look at this theory on an ordinary continuum $2-$sphere.

Let $\phi$  be a real scalar  field on a  sphere $\mathbb{S}^{2}$ with
radius $R$ and $\phi^{4}$ potential, the functional action is given as 
\begin{equation}\label{eq:theactionX}
  S  \left[ \phi  \right]  = \int_{\mathbb{S}^{2}}  \ud \Omega\,  \left[
  \frac{1}{2} \left(  \nabla \phi \right)^{2} +  \frac{1}{2}rR^2
 \phi^{2} + \frac{1}{4!} \lambda  R^2 \phi^{ 4 }
  \right],  
\end{equation}
where $\nabla_{i}$ ($i=1,2,3$) are  the usual generators of rotations,
$r$ is the  \emph{mass parameter}\index{Mass parameter} and $\lambda$
is  the \emph{coupling  constant}\index{Coupling constant}  which may
depend on the radius of the sphere.  

Second  order phase  transitions can not  appear in  finite volume
systems, such as the sphere. However, it becomes possible in the
\emph{planar limit}\index{Planar!limit}, $R \to \infty$.  
The $\phi^{4}$  model on  a bidimensional plane, which corresponds to
the planar limit  of the sphere, is defined by the action 
\begin{equation}\label{eq:theactionplane}
  S  \left[ \phi  \right] =  \int_{\mathbb{R}^{2}}  \ud^{2} \mathbf{x}
  \left[  \frac{1}{2} \left(  \nabla \phi  \right)^{2}  + \frac{1}{2}r
  \phi^{2} + \frac{1}{4!} \lambda \phi^{ 4 } \right].  \nn 
\end{equation} 
This  model  has  been  widely  studied, see for
example~\cite{pending1,pending2}.

Similarly, the model to study on the fuzzy sphere is a \emph{Hermitian
  matrix model}\index{Hermitian matrix model} which corresponds to a
  real scalar field and is given by the action \cite{Martin:2004un,den}
\begin{equation}\label{eq:accion2}
  S \left[ \phi ; N,a,b,c \right] = \traza{  a \left[ \hat{L}_{i} , \phi
  \right]^{\dagger} \left[ \hat{L}_{i} , \phi \right] + b \phi^{2} + c
  \phi^{4} }= \traza{  a  \phi^{\dagger} (\mathcal{L}^2 \phi) + b \phi^{2} + c
  \phi^{4} },
\end{equation}
where $N$ is the size of the matrix, $b$ is the real \emph{mass
  parameter}\index{Mass parameter}, and $c$ is the real, positive,
\emph{coupling constant}\index{Coupling constant}. Similarly to a
commutative sphere, $\left[ \hat{L}_{i}, \cdot \right]$ are the usual
rotation generators where $\hat{L}_{i}$ is the \emph{angular momentum
  operator}\index{Angular momentum operator} in its irreducible
representation of $\su{2}$ with size $N = \left( 2 \ell + 1 \right)$
defined by the commutation relations
$[\hat{L}_i,\hat{L}_j]=\varepsilon_{ijk}\hat{L}_k$. The constant $a$
is a positive number employed to fix the units\footnote{It is possible
  to scale $\phi$, $b$ and $c$ to absorb $a$ \emph{i.e.} fit $a$ to
  one. The scaling for the field is given by: $\phi= \psi/\sqrt{a}$,
  leading to a scaling for the other parameters of $\tilde{b} 
  =b/a$, and $\tilde{c}=c/a^{2}$. These changes affect the expectation
  values by a constant overall scaling which has no consequence on
  such things as phases and phase boundary
  lines.\label{foot:scaling}}.  The $a$ term, called \emph{kinetic
  term}\index{Kinetic term}, contains the information on the geometry
of the space by means of the Laplacian, while the rest of the action
is called \emph{potential term}\index{Potential term} and denoted
$V(\phi)$.

The action (\ref{eq:accion2}) approximates the continuum action
(\ref{eq:theactionX}) when
\begin{equation}\label{eq:limitaccion}
a=\frac{2\pi}{N},\ b=\frac{2\pi rR^2}{N},\ c=\frac{\pi\lambda R^2}{6N}.
\end{equation}
These parameters are chosen so that the fuzzy action was normalised so
that $S [ \hat{1} ]$ for the unit function/matrix be the same on the
continuum and fuzzy sphere.

The absolute minima of this action can be obtained by searching for
configurations minimizing both the kinetic and potential term {\it
  separately}. The kinetic term is obviously positive and is therefore
minimum when $\mathcal{L}^2\phi=0$, that is when $\phi = \alpha
\hat{\mathbf{1}} $ is proportional to the identity. Replacing
this constraint in the potential term, we get
\begin{equation}\label{eq:classicalminimum}
  V \left( \alpha \hat{\mathbf{1}}\right) = N(b \alpha^{2}+ c \alpha^{4}).  
\end{equation}
The necessary
conditions to have a minimum are: $ S^{ \prime
} \left( \overline{\alpha} \right) =0$ and $S^{ \prime \prime }  \left(
\overline{ \alpha } \right) > 0 $. If $b<0$ then
we find two absolute minima at

\begin{equation}\label{eq:inn}
  \overline{\alpha} = \pm \alpha_{_0}= \pm \sqrt{ \frac{- b }{ 2 c } }, 
\end{equation}
which have energy 
$S \left( \overline{\alpha} \right) = - N b^{2}/4 c$,
whereas when $b>0$ there is only one minimum at $ \overline{ \alpha } = 0 $. 
Finally, when $b=0$, there is a critical point at $ \overline{ \alpha } = 0 $
which is clearly a minimum since $S(\alpha)=Nc\alpha^4$ and $c>0$.

There are however other {\it local minima} to this action which will
play an important role in one of the phases of this model. They can be
located approximately by looking at the minima of the potential
\cite{Martin:2004un} which are given by $U^\dag DU$ where $U$ is a
unitary matrix and $D$ a diagonal matrix with diagonal elements $\pm
\alpha_0$. The absolute minima found above correspond to the
particular case when all the diagonal entries of $D$ are identical.

 \section{The Metropolis simulation}\label{sec:Metropolisalgorithm}
 We started the simulations by using a standard Metropolis Monte Carlo
 algorithm \cite{Metropolis,Metropolisbooks} with the jackknife
 estimator for the error \cite{shaotubook} to account for the
 autocorrelation of the samples..

The \emph{initial conditions}, {\it i.e.} the choice of the first
configuration in the Markov chain can be of two types: \emph{Cold
initial conditions}, \index{Initial conditions!cold} which correspond
to configurations which are classical minima of the action, or
\emph{Hot initial conditions}\index{Initial conditions!hot}, which are
configurations chosen randomly in the phase space. We made sure in our
numerical simulations that none of our results depended on the initial
conditions, whether they were cold or hot.

In general, when we start the simulation the sequence of samples
obtained by Metropolis algorithm goes through a transient regime where
it does not obey the desired statistics yet. This is the
\emph{thermalization process}. This is true even in the case of
``cold'' initial conditions because the classical
minima may be probabilistically irrelevant when the fluctuations are
important. This actually happens in one of the phases of our model
(the non-uniform ordered phase).

\emph{Tunneling} is an essential process in our model as there are
multiple classical minima which contribute significantly to the
probability distribution of the field. Typically, tunneling is
exponentially suppressed by the energy barrier separating the
classical minima. It can therefore be difficult to account for in the
Monte-Carlo algorithm. To improve the probability of tunneling, we
have tried various sampling methods.

The two simplest ways of sampling the phase space are to either make a
big change on the matrix as a whole or to perturb its entries one by
one. The first method allows for big changes and helps tunneling but
usually yields unfavored, high energy, test configurations which are
rejected and increase the autocorrelation between configurations. On
the other hand, the latter is good at exploring the phase space
locally, but has a low chance of tunneling. Even alternating the two
methods to enjoy both their advantages is not sufficient to produce
the tunneling necessary in the model studied.

As we already discussed at the end of
section~\ref{sec:modelonafuzzysphere}, the classical minima of the
action are located at $\pm\sqrt{-b/2c}$. Thus, the interval where we
must vary the real and imaginary part of every entry of the matrix
during the sampling should be about $I=[-2\sqrt{-b/2c};2\sqrt{-b/2c}]$.
In practice, we have found empirically that we need an interval of
variation of the field between $2.3$ and $2.6$ times bigger.

When we use an interval less than $2.3\,I$, the trace effective
probability density distributions of the matrix will not reproduce the
results obtained via direct integration for $N=2$. In general,
this effect also appears for any matrix size $N$. The upper bound does
not affect the results so much as the auto-correlation of samples
(more configurations are rejected by the metropolis algorithm) and
thus the speed of convergence of the code.  We have found that $2.6I$
is the optimum upper bound to balance speed and precision.

A more sophisticated method we have successfully implemented is the
\emph{annealing method} \cite{Metropolisbooks}. The idea is to produce
favored decorrelated test configurations by introducing a
temperature-like parameter $\beta$ in the probability distribution
$\exp(-\beta S)$. The Metropolis sampling is done normally with
$\beta=1$. During that sampling, the field is typically trapped around
one of the classical minima. Periodically, the Metropolis sampling is
interrupted, and this parameter is lowered ({{\it i.e.} the
temperature is increased) which smoothes out the action and allows the
field to move more freely between the classical minima. Then $\beta$
is raised back to one ({\it i.e.} the temperature is lowered back)
trapping the field around a classical minimum which is hopefully
decorrelated from the previous one.

The annealing method thus increases the probability of tunneling
between minima of the action and decreases the autocorrelation between
configurations. It also increases the computation time, but the
gain in efficiency is largely dominant, making this method 
very useful and reliable for simulations with larger matrices.

The computation time can still be too large, making the simulation
impossible to run in practice. We have developed a method where
the real time of computation decreases dramatically which we
present in the appendix.

\section{Limiting models}
In this section we present the lowest dimensional model which can be
integrated directly and the pure potential model which can be solved
analytically. They will both be useful to test the validity of our
Monte-Carlo simulation in some particular limits. 

\subsection{Lowest dimensional model}\label{sec:lowest}
It is quite useful to investigate the lowest dimensional model ($N=2$)
as it has only two independent parameters and can therefore be well
understood, and integrated directly. This provides a good independent
computation to test our Monte-Carlo code against. Furthermore, it
happens that even this low dimensional case shows all the features of
the large $N$ limit!

In  the simplest  case  $N=2$, the  action  (\ref{eq:accion2}) can  be
simplified by  expanding the field  in terms of an orthonormal basis
$\{ {\bf  1} ,  \sigma_{k} \}$, where  ${\bf 1}$  is the $2  \times 2$
identity   matrix   and  $\sigma_{k}$   are   the  three   \emph{Pauli
  matrices}. The expansion is 
\begin{equation}\label{eq:theexpansion2x2}
  \phi = \alpha {\bf 1} + \vector{ \rho} \cdot \vector{\sigma}
\end{equation}
where   the   coefficients   $\alpha$   and  $\rho_{k}$   are   in   $
\mathbb{R}$. Then, writing down the action (\ref{eq:accion2}) in terms
of this new set of variables, we get 
\begin{equation}\label{eq:theaction2x2}
S = 4 a\rho^{ 2 } + 2 b \left( \alpha^{ 2 } + \rho^{ 2 } \right) + 2 c
\left( \alpha^{4} + 6 \alpha^{2} \rho^{2} + \rho^{4} \right) 
\end{equation}
where    $\rho$ is the norm of $\vector{ \rho } $.  

The action (\ref{eq:theaction2x2}) depends only on the modulus of the
  vector $\vector{ \rho}$.  This property allows us to integrate out the
  degrees of freedom associated with the \emph{rotational
  symmetry}\index{Rotational symmetry} of the vector $\vector{\rho}$, which is
  the expression of the general $SU(2)$ symmetry of the action in two
  dimensions. The corresponding effective probability density distribution is
  given by
\begin{equation}\label{eq:probability2x2}
  P_{\mathrm{eff}} \left[ \alpha , \rho \right] = \frac{1}{Z}
    \rho^{2} \ex^{-S \left[ \alpha , \rho \right] }= \frac{1}{Z}
    \ex^{-S_{\mathrm{eff}} \left[ \alpha , \rho \right] }, \qquad Z =
    \int \ud \alpha \ud \rho \rho^{2} \ex^{-S}= \int \ud \alpha \ud
    \rho \ex^{-S_{\mathrm{eff}}},
\end{equation}
where $ S_{{\mathrm{eff}}} =S - \ln \rho^{2}$ is the associated
effective action.

This  simple example  depends on  two  variables only,  which makes  it
possible to integrate numerically without  a lot of effort for any set
of  parameters  $\{a,b,c\}$ via  the  trapezoidal  rule  or any  other
algorithm  \cite{numerical}, to  get  the expectation  values. In  this
sense, \emph{we can solve directly the model for any set of parameters
  making it possible to test our  Monte Carlo code}. Any graph in this
section will not include  error bars because, with direct integration,
they are negligible.  

\begin{figure}[tbp]
  \begin{center}
\plotone{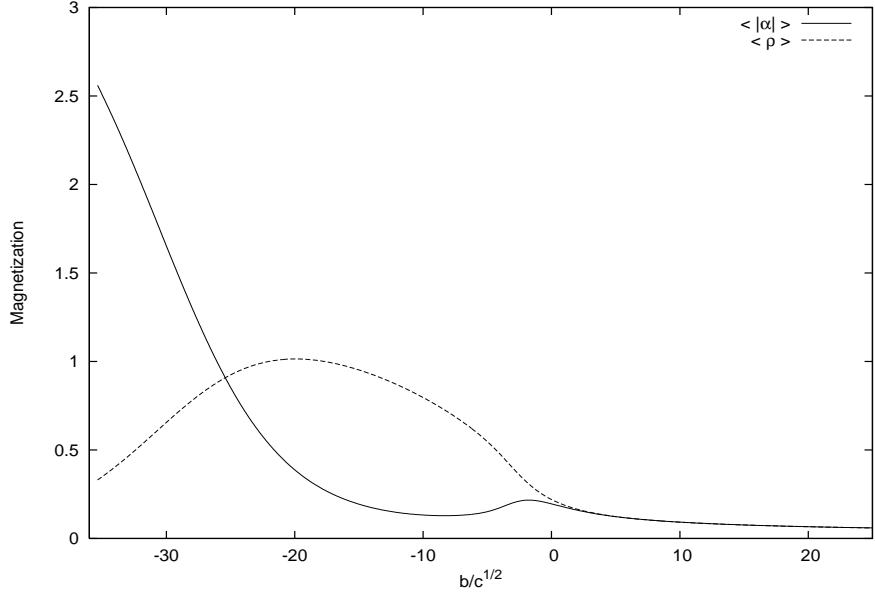}
\end{center}\caption{Expectation value of $|\alpha|$ and
  $\rho$ obtained from direct numerical integration.}\label{fig:magnet_2x2}  
\end{figure}

Because of the $SU(2)$ symmetry of the theory, the expectation values
of $\left< |\alpha| \right>$ and $\left< \rho \right>$ give us the
whole information about the average configuration $\left< \phi
\right>$. We can see their behavior in the figure
\ref{fig:magnet_2x2}, computed from a direct integration with $N=2$,
$a=1$, and for a typical value of $c=50$, as a function of the
remaining parameter $b$ of the model (scaled to $ b c^{-1/2}$). We can
see three distinct phases: 
\begin{enumerate}
\item   \emph{Disordered  phase}:\index{Phase!disordered}   the
  expectation values  of $\left| \alpha \right|$ and  $\rho$ are close
  to zero, roughly in the interval $\left(  0 ,  + \infty  \right)$. 
\item \emph{Uniform ordered phase}:\index{Phase!uniform ordered} the
  most important contribution to the configuration is given by the
  expectation value of $\alpha$, roughly in the interval $\left( - \infty
  , -24 \right)$
\item \emph{Non-uniform ordered phase}: \index{Phase!non-uniform
ordered} the system is ordered but the main contribution to the
configuration is given by the expectation value of $\rho$, in between the
previous two phases.
\end{enumerate}

The disordered\index{Phase!disordered} phase has averages of
$|\alpha|$ and $ \rho$, in the expansion (\ref{eq:theexpansion2x2}),
at approximately $0$ and the typical configuration is thus distributed
around zero.  This phase is analogous to the \emph{paramagnetic
  phase}\index{Phase!paramagnetic} in ferromagnetic materials.
Following the analogy, if we take the parameter $b$ as a
``\emph{temperature parameter}''\index{Temperature parameter}, the
thermal fluctuations do not allow any kind of ordering in the material
when the temperature is bigger than some critical value. The thermal
fluctuations are getting stronger and stronger when the temperature is
increased.

The uniform ordered\index{Phase!uniform ordered} phase is characterized
by the  fact   that  the   most  important   contribution   to  the
configurations is given by  the coefficient $\alpha$, in the expansion
(\ref{eq:theexpansion2x2}).  The   expectation  value  of   $\rho$  is
negligible with  respect to the expectation value  of $|\alpha|$. This
means  that the  configuration  is approximately  proportional to  the
identity  matrix.   This  phase   is  analog  to   the  \emph{magnetic
  phase}\index{Phase!magnetic}, in ferromagnetic materials.  

In the third and last phase (the non-uniform ordered
phase)\index{Phase!non-uniform ordered}, both $|\alpha|$ and $\rho$
contribute to the configuration but in this region of the parameter
space, $\rho$ is more important than $|\alpha|$.  This phase has
ordering \emph{ i.e.} the field has non-zero expectation value but
this ordering is not an analog of any ferromagnetic ordering.  It was
argued in \cite{marco,Panero:2006cs} that in this phase, the
eigenvalues of the matrix has two cuts located at the two minima of
the action $\pm\alpha_0$ given in (\ref{eq:inn}), whereas
\cite{Martin:2004un} speculated that the eigenvalues of $\phi$ would
be split equally between positive and negative eigenvalues. For $N=2$,
it means trivially that $\phi=\alpha_0\sigma_3$ up to a free $\su{2}$
rotation, and thus one would expect $<|\alpha|>\ll \alpha_0$ and
$<\rho>\simeq \alpha_0=\sqrt{-b/c^{0.5}}/(4c)^{0.25}$. This is indeed
true in figure \ref{fig:magnet_2x2}, as for $-20<b/\sqrt{c}<-5$,
$<\rho>$ does curve like $\sqrt{-b/c^{0.5}}/200^{0.25}\simeq
\sqrt{-b/c^{0.5}}/4$ and is much bigger than $<|\alpha|>$.

As stated earlier, we can also use the results from this alternate
method to validate the Monte-Carlo code for $N=2$.  The
figure~\ref{fig:probN2a1MCcompara} shows the unnormalised effective
probability density distributions of the quantities $\alpha$ and
$\rho$.  In that case, we can compare directly both effective
probability density distributions.  The excellent agreement shows that
the statistical error bars are negligible, but the most important
thing is that the program has sampled the phase space properly.
We have already checked many points in the parameter space and we have
always obtained identical results up to error bars.

\begin{figure}[tbp]
  \begin{center}
  \subfigure[Probability distribution of $\alpha$.]{\scalebox{0.92}{
  \label{fig:PDDalpha}  
\plottwo{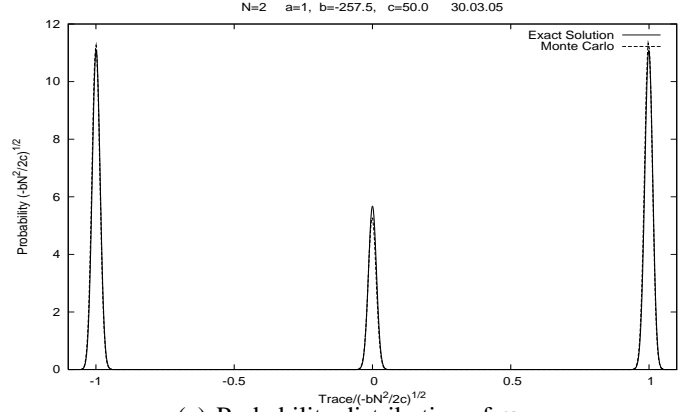}}} \quad
  \subfigure[Probability distribution of $\rho$.]{\scalebox{0.92}{
  \label{fig:PDDrho} 
\plottwo{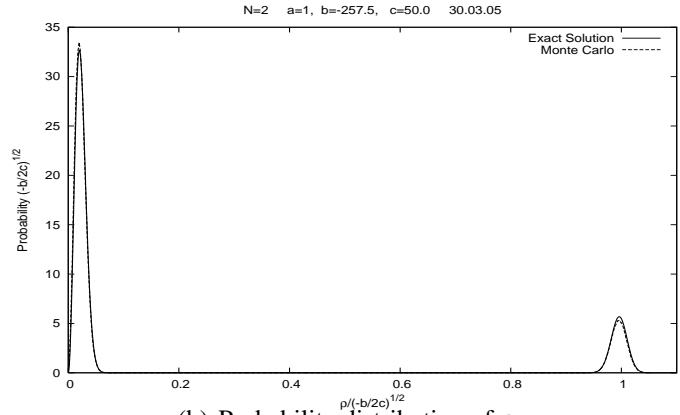}}}
\caption{Comparison of the unnormalised probability density
  distributions of some observables for $N=2$. In most cases, the two
  curves cannot be distinguished.}\label{fig:probN2a1MCcompara}
\end{center}
\end{figure}

At the moment, we have shown the convergence of our simulation in the
\emph{lowest dimensional model}\index{Lowest~dimensional~model}, but
our goal is of course focused on simulating the model using bigger
matrix sizes to extrapolate to the \emph{continuum
limit}\index{Continuum limit} $ \left( N \rightarrow + \infty
\right)$. Still we will see that the $N=2$ results are already
remarkably good approximations of the large $N$ limit.

\subsection{Pure potential model}\label{sec:purepot}

The \emph{pure potential model} interests us for two reasons: it can
be solved analytically and gives a good approximation for the
transition curve between the disordered and non-uniform ordered phases
discussed previously \cite{Martin:2004un}.  It comes from setting
$a=0$ in the action (\ref{eq:accion2}), only keeping what we called
the potential term. This approximation is increasingly accurate as the
transition is tracked to larger couplings far from the triple point.

This model ($a=0, N \to \infty$) has been solved by many authors
\cite{Shimamune:1981qf,pavel}. In term of their solution we can get an
expression of the specific heat and other thermodynamics quantities
which are a good reference to compare to the numerical results and the
convergence of the algorithm when we increase $N$.

The specific heat in this approximation has the form
\begin{equation}\label{eq:Cv_exact}
  C_{V} = \left\{ \begin{array}{lcl}
                  \frac{1}{4} & \null & \overline{r} < -1   \\ 
                  \frac{1}{4} + \frac{2 \overline{r}^{4}}{27} -
                  \frac{\overline{r}}{ 27}
                  \left( 2 \overline{r}^{2} - 3 \right) \sqrt{
                  \overline{r}^{2} + 3} & \null 
                   & \overline{r} >-1
\end{array}
\right.
\end{equation}
where $\overline{r}=b/|b_{c}|$ with $b_{c}=-2\sqrt{ Nc }$ is the
critical mass. From equation (\ref{eq:Cv_exact}) the phase transition
is a third order transition because the first derivative of the
specific heat has a finite discontinuity in $\overline{b} = -1$.

Another way to detect the phase transition is to look at the
probability distribution of the field eigenvalues. In the disordered
phase, they are confined into a single connected region centered
around zero, whereas in the non-uniform ordered phase, they are split
into two disconnected regions centered respectively around $\pm
\sqrt{-b/2c}$ corresponding to the minima of the polynomial
potential. Due to this characteristic behaviour, we also refer to this
as a ``one cut--two cut'' transition.  We will also use this
terminology for the disorder/non-uniform transition since the work of
Panero \cite{marco,Panero:2006cs} shows that the transition in the
fuzzy sphere model, where $a\neq0$, also have this characteristic
behaviour.

The phase boundary for this model is given by \begin{equation} 
b=-2\sqrt{Nc} \label{eq:pbppm} \end{equation}
and is included in the phase diagram shown in
figure \ref{fig:thephasediagram} at the end of this article.

\section{Observables and Simulations}\label{chap:simulation}
For the model under study, the number of degrees of
freedom\index{Degree of freedom} is $N^{2}$, which corresponds to the
number of independent real entries in a Hermitian matrix.  Thus, the
thermodynamic limit we are interested in corresponds to matrices of
infinite size. The standard procedure, to take the thermodynamic
limit, is to define a scaling of the parameters of the model such that
the relevant observables collapse in a phase diagram independent of
the matrix size. If the phase diagram collapses in a reasonable way,
then we can straightforwardly extrapolate it to the thermodynamic
limit}.

\begin{figure}[tbp]
  \begin{center}
\plotone{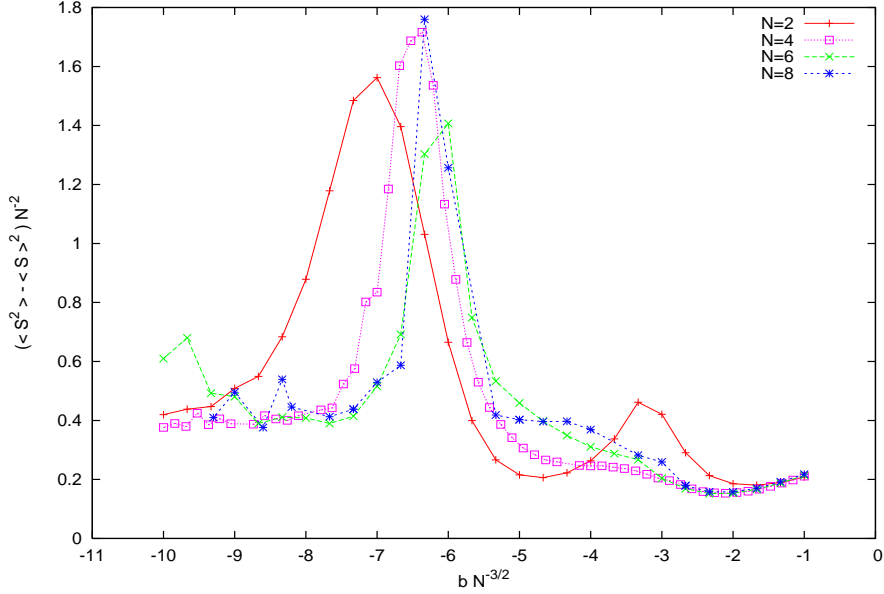}
\end{center} \caption{Specific heat for $ a=1 $ and
  $c=40$.}\label{fig:specificheat}
\end{figure}

The specific heat is a measure of the dispersion of the energy. It is
sensitive to the phase transitions which register as peaks in it.  We
therefore use it as the order parameter. Typically, it will present
one or, more often, two peaks as we show in
figure~\ref{fig:specificheat} for $\{ a=1$, $c=40\}$ and various
matrix sizes.  The very obvious peak is located around $bN^{-3/2}=-6
\pm 0.4$, the other one, almost imperceptible is around
$bN^{-3/2}=-3$. For the biggest matrix size investigated $N=10$,
simulations for a curve as the one in figure \ref{fig:specificheat}
took about a day. The error bars provided by the jackknife algorithm
were omitted as they are quite small and would only crowd the figure
more.

Other observables, such as $<\traza{\phi^2}>$ and $<|\traza{\phi}|>$
which were used as order parameters in \cite{Martin:2004un}, their
susceptibilities, and the internal energy $<S>$ have also been collected
but are not shown here. They will be used in section
\ref{chap:results} to identify the phases though.

It is an important remark that the transition, from the non-uniform
ordered phase to the uniform ordered phase, presents a very high and
wide peak in the susceptibilities which can subsume and hide the
smaller one when near the triple point, making it impossible to
determine its exact position. As a result, the data points of this
transition curve in the phase diagram could not be found near the
triple point. However, Panero \cite{marco}, by looking at the
eigenvalue distribution of $\phi$ for $c/aN^2=1/2$ provides an
additional point on this curve very near the triple point.

In the figure \ref{fig:specificheat}, the scaling for $b$, given by
$bN^{-3/2}$, which aligns the peaks (and thus the location of the
phase boundary) for different matrix sizes has already been included.
It is remarkable that with this scaling, the $N=2$ curve has the same
qualitative behavior as the $N=10$ curve, as announced previously, but
the peak in figure \ref{fig:specificheat}, is already a reasonable
approximation of the large $N$ limit peak found for $N=10$.

This analysis was repeated for a wide range of the parameter $c$ and
for matrix sizes $N\leq 10$. The collected results, the interpretation
of the phases and the collapsed phase diagram will be presented in
section \ref{chap:results}.

\section{Results}\label{chap:results}
In this section,we will present the collapsed phase diagram as well as
an analysis of the three phases observed.

\begin{figure}[tbp]
  \begin{center} 
\plottwo{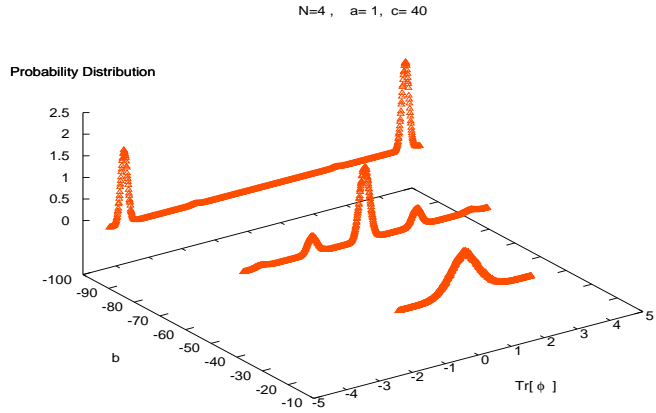}
  \end{center}  \caption{Probability distribution  of the  trace  of $
    \phi $ as a function of $b$}\label{fig:alpha} 
\end{figure}

\begin{figure}[tbp]
  \begin{center} 
\plottwo{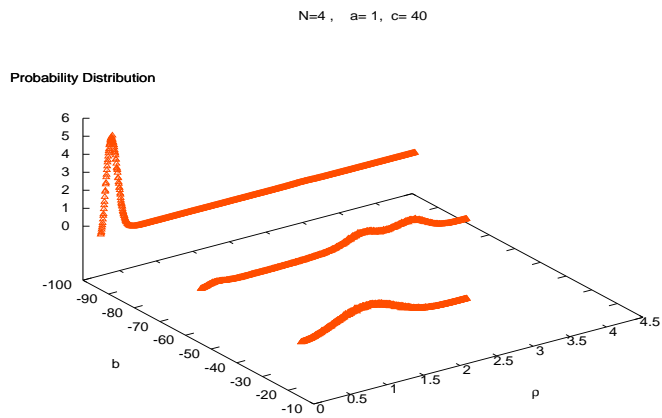}
  \end{center}\caption{Probability           distribution           of
    $\rho$ as a function of $b$.}\label{fig:rho} 
\end{figure}

In the plots \ref{fig:alpha} and \ref{fig:rho}, we can see different
profiles of the probability density distributions as a function of the
mass parameter $b$ for $\traza{\phi }$ and $\rho=\sqrt{
|c_{_{1\ -1}}|^{2} + | c_{_{10}}|^{2} + |c_{_{11}}|^{2} }$ which gives
the power in the $l=1$ angular momentum mode in
(\ref{eq:fuzzyexpansion}), with $\{ a=1 $, $c=40$, $N=4 \}$. There, we can appreciate the three different phases
of the model.


In the uniform order phase, for $b$ negative enough, the trace is
distributed around two symmetric values centered on $\pm \alpha_{0}$
respectively, \footnote{$\alpha_{_0}$ was defined in (\ref{eq:inn}) as
the location of the absolute minimum of the action.}, and $\rho$ is
distributed close to zero, {\it i.e.} it gives no contribution to the
typical configuration. In this phase, $\phi$ is approximately
proportional to the unit matrix and the rotational symmetry is thus
preserved.

The non-uniform ordered phase, for intermediate values of $b$, has the
peculiarity that the most exterior peaks of the probability of the
trace, which correspond to the absolute minimum of the action $\pm
\alpha_{0}$ and thus to the field in the uniform order phase, are
smaller than the new peaks which arise between them. Furthermore, the
most probable value of $\rho$ is not close to zero. In this phase, the
power of the configuration is thus in higher angular momentum modes
(as defined in the expansion (\ref{eq:fuzzyexpansion}) in polarization
tensors) and the rotational symmetry has been spontaneously broken.

The last curve is representative of the disordered phase. In this
phase the configurations (both the trace and $\rho$) are spread over a
long interval but very close to zero restoring the rotational
symmetry.

\begin{figure}[tbp]
  \begin{center} 
  \subfigure{\scalebox{0.92}{\plotone{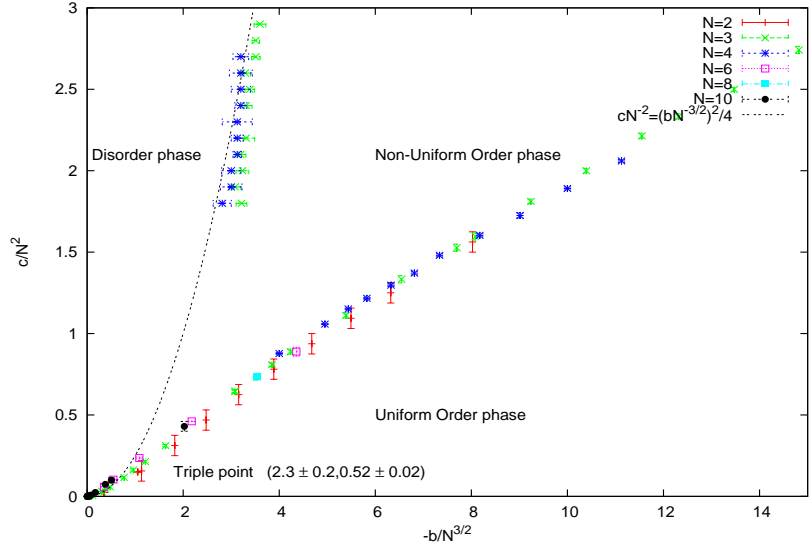}}}
\caption{Full collapsed phase diagram.} \label{fig:thephasediagram} 
\end{center}
\end{figure}

A phase diagram is a map that contains the thermodynamics or physical
properties of a given system.  This implies that, to construct a phase
diagram, we need quantities in the thermodynamic limit. As explained
in section \ref{chap:simulation}, this is done by finding a scaling in
the bare parameters of the model, $b$ and $c$ here, to make it
independent of the number of degrees of freedom $N$. 

We had already found in section \ref{chap:simulation} that the scaling
necessary to make the diagram independent of $N$ was
$N^{-\frac{3}{2}}$ for the mass parameter $b$. Repeating the
simulations for various values of $c$ and plotting the phase boundaries
found for all values of $N$ simulated, we found a scaling in $N^{-2}$
for the coupling constant $c$. This scaling is the same for \emph{all}
the coexistence curves which guarantees a consistent $N \to \infty$
limit. We can then define scale--free parameters \begin{equation}
\bar{b}=\frac{b}{aN^{3/2}}, \ \bar{c}=\frac{c}{a^2N^{2}}.\label{bcbar}
\end{equation}
Remember that for all the simulations and results in this paper, we
have set $a=1$. 

These results are presented in the figure \ref{fig:thephasediagram}
which shows the phase diagram for the $\phi^{4}$ model on a fuzzy
sphere. The three phases we identified above are delimited by the
coexistence curves which meet at a triple point.  These coexistence
curves can be fitted to get an algebraic expression for each one of
them using the scale-free parameters introduced above in (\ref{bcbar})

As mentioned in section \ref{chap:simulation} we could not access the
Disorder/non-uniform order phase boundary near the triple point.
However since the curve is consistent with a straight line, we can
extrapolate it to the triple point without any difficulty.  
We find:
{\blue
\begin{equation} \mbox{Disorder/non-uniform order: }
\bar{c} = 2.29 (-\bar{b}) - 4.74 .\label{eq:dis-nuo2}
\end{equation}}
As expected, for large ${\bar c}$, this curve is well approximated by
the one obtained for the pure potential model derived in
\ref{sec:purepot}, given in scale--free parameters by
\begin{equation} \bar{c}=(-\bar{b})^2/4,\label{eq:ppotmodred} \end{equation}
and drawn with a dashed line in the phase diagram, figure 
\ref{fig:thephasediagram}. 

We did not focus on the disorder/uniform order boundary line in this
paper since it has already been studied in detail in
\cite{Martin:2004un}. It was found there to be a straight line going
through the origin. Converting its equation to our scale--free
parameters through (\ref{bcbar}), we get 
{\blue
\begin{equation} \mbox{Disorder/uniform order: } 
\bar{c}=0.23 (-\bar{b}).\label{eq:dis-uo}
\end{equation} }

Finally, the uniform--non-uniform order phase boundary line which was
studied in detail in this paper is approximately straight with
equation
{\blue
\begin{equation} \mbox{Uniform - non-uniform order: }
\bar{c}=0.2 (-\bar{b})+0.07 \label{eq:u-nuo}
\end{equation}}
which just prolongs the disorder/uniform order line, up to error bars.

These three coexistence curves,
(\ref{eq:dis-nuo2},\ref{eq:dis-uo},\ref{eq:u-nuo}), intersect at
a triple point given by
\begin{equation}
(\bar{b}_T,\bar{c}_T)=(-2.3, 0.52). 
\end{equation}
These values are consistent with the data presented in \cite{marco}.
In fact, figures (11-30) there correspond precisely to
$\bar{c}=0.5\simeq \bar{c}_T$, and by identifying the point where the
eigenvalue density undergoes the one cut--two cut transition described
in section \ref{sec:purepot}, one finds that his data gives
$\bar{b}_T\simeq 2.3$ consistently for $N=15,17,19,21,23$.

If instead one takes the asymptotic form of the disorder non-uniform order
transition line given by the one cut--twocut transition
(\ref{eq:ppotmodred}) instead of (\ref{eq:dis-nuo2}), and finds its
intersection with the disorder--uniform order transition curve
(\ref{eq:dis-uo}), the triple point occurs at
\begin{equation}
(\bar{b}^e_T,\bar{c}^e_T)=(-0.92,0.21).\label{bceT}
\end{equation}
We conclude from this that the effect of the kinetic term is to move the triple
point to larger values along the line governing the disorder/uniform order transition.

\section{Conclusions}\label{chap:conclusions}
In the main part of the paper, we presented the results for the
numerical simulation by means of an optimized Metropolis algorithm for
the $\phi^{4}$ matrix model. In the appendix we develop the metropolis
algorithm which makes more efficient the simulation of matrix models.
In particular, we argue that the algorithm proposed presents
considerable advantages with respect to the usual Metropolis algorithm
in the simulation of matrix models \cite{tesis:fergar}. The reduction
in the processing time for both non-uniform ordered and uniform
ordered phases will be more evident for large matrices and, of course,
when we are far away from the coexistence curves due to the fact that
the minima of the potential are more separated. A different approach
was used with equal success in \cite{marco}.

Figure \ref{fig:thephasediagram} shows the phase diagram for the model
given by (\ref{eq:accion2}) and refines the phase diagram which was
incomplete in \cite{Martin:2004un}. The data have been collapsed using
the scaling form shown on the axis and defined in (\ref{bcbar}). It is
consistent with the scaling of the exact solution of the pure matrix
model which only fixed the quotient of the two scalings.  One of the
important features of the diagram is that all three coexistence lines
can be collapsed simultaneously. This did not have to happen and in
fact the corresponding lines do not all collapse together for a
related three dimensional model \cite{Medina:2007nv}, where the
spacetime is taken to be a fuzzy sphere direct producted with a
temporal direction.

This diagram contains the information about three different phases,
the well known disordered and uniform ordered phase, and a new phase,
the non-uniformly ordered phase (where the $SO(3)$ spatial symmetry of
the round sphere is spontaneously broken), as well as the scaling of
the model, and the coexistence curves.  We could even estimate the
coordinates of the triple point which is the point where the three
phases coexist in equilibrium. The coordinates of this triple point
are consistent with the independent simulation \cite{marco}.

Another article \cite{Das:2007gm} finds different results, including
an extra phase and no scaling. An obvious reason may be that in the
phase diagram they show for $N=25$, our scale--free parameter
$-\bar{b}$ has a very small range in our scale--free parameters of
$[0;0.13]$, meaning it only shows a tiny corner of our phase diagram
of figure \ref{fig:thephasediagram}. Furthermore, they use the
probability distribution of $\phi_{11}$ (denoted $\Phi_{11}$ there) as
an observable to detect the transition between the two ordered
phases. First, this does not seem to be a physically meaningful
observable, especially given the $SU(2)$ symmetry of the
model. Furthermore, they obtain a curve somewhat similar to the one
cut curve for the eigenvalues of $\phi$, but they locate the
transition when this profile gets deformed with a dip at zero, instead
of when it switches to two cut (if it ever does).  This boundary line
is absent in our phase diagram and in previous ones
\cite{Martin:2004un,marco,Panero:2006cs}, and we find no evidence for
such a transition or a new phase in this region of the phase diagram.
As for the lack of scaling for the other phase boundaries, it is
difficult to decide the cause, but it is disquieting that their
simulations sometimes depend on the initial condition, such as when
they find hysteresis.

In the large $N$ limit the model with $a=0$ has a third order phase
transition between disordered and non--uniform ordered
phases \cite{Shimamune:1981qf,pavel}. The disordered phase
is described by a single connected eigenvalue distribution called a
``one cut phase'' distribution, whereas the ordered phase is
described by an eigenvalue distribution split into two disconnected
distributions centered on opposite values and called a ``two cut
phase'' distribution. The transition occurs when the two cuts merge to
become a single cut for $c=b^{2}/4N$.
\begin{figure}[!h]
\begin{center}
\includegraphics[width=0.53 \textwidth,angle=-90]{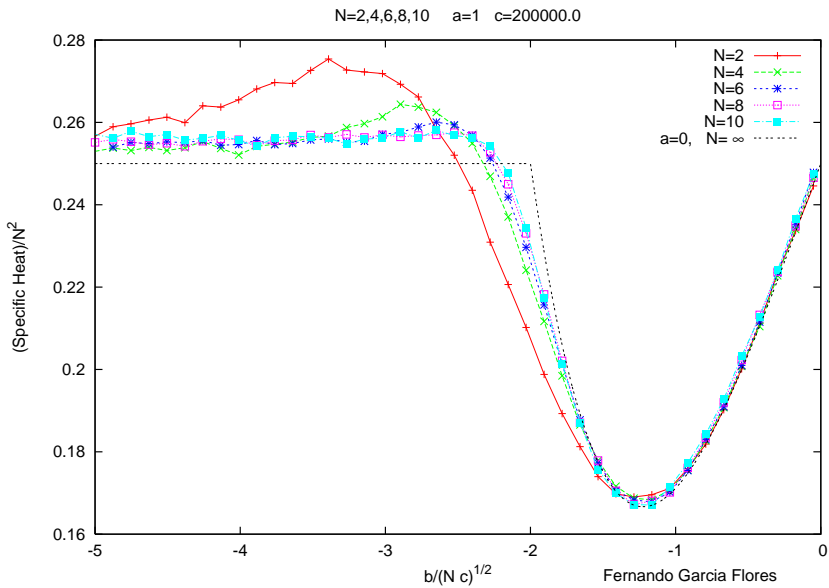}
\caption{Plot of the specific heat at the disordered/non--uniform
  ordered transition for increasing $N$ and its $N\to\infty$
  limit, the exact pure potential model, given by Eq. 
(\protect\ref{eq:Cv_exact}).}
\label{fig:matrix}
\end{center}
\end{figure}
Figure \ref{fig:matrix} confirms numerically the
convergence of the disordered/non--uniform ordered transition towards
this exact critical line of the pure potential model as the coupling is
increased. The simulations of Panero \cite{marco,Panero:2006cs} confirm that
this transition for the full model is indeed a one cut--two cut transition
though the eigenvalue distribution now has a richer structure.

We expect that the existence of the cut transition of matrix models
and of a triple point is a generic feature of fuzzy scalar field
models, since all such models should reduce to a pure matrix model
when the kinetic term becomes subdominant. This means that fuzzy
scalar field models should generically have an exotic phase with
spontaneously broken spacetime symmetry.

Numerically, it is not difficult to find the coexistence curve between
the uniform ordered and disordered phases which exist for low values
of $c$. On the other hand, the coexistence curve between the two
ordered phases is difficult to evaluate because it involves a jump in
the field configuration and tunneling over a wide potential barrier.

In the current model the triple point is estimated to be located at
\begin{equation}
(\bar{b}_T,\bar{c}_T)=(-2.3,0.52).
\end{equation}
This is obtained by extrapolating the three coexistence lines till
they meet. Surprisingly good agreement with this result was obtained
in \cite{O'Connor:2007ea} by performing perturbation theory in the
kinetic term, {\it i.e.}  by expanding in the parameter $a$ to second order.
It is not however, totally clear that the triple point identified
there coincides with the one here as a different scaling of the
parameters was necessary, but the salient features are the same.

The position (\ref{bceT}) where the curve (\ref{eq:ppotmodred})
intersects the disordered /uniform order transition line suggests
that the effect of the kinetic term is to push the transition, and
hence the triple point, to larger negative values of $b$. This is a
positive feature since it indicates that adding a higher derivative
term to the model will allow one to tune the triple point to large
coupling. The conjecture is that this will be sufficient to eliminate
the UV/IR mixing problems \cite{den} and recover the commutative
theory with the correct fluctuations \cite{DOP}.

It still remains to be seen what thermodynamic limits can be drawn
from the phase diagram and the scalings in each of the limits of the
fuzzy sphere introduced in table \ref{tab:fuzzyspaces}: the
ordinary sphere, the ordinary plane, and the Moyal plane.

To that end, we want to reexpress the positions of the coexistence
curves
(\ref{eq:dis-nuo2},\ref{eq:dis-uo},\ref{eq:u-nuo})
which depend on $a=1$, $b$, $c$ and $N$ as a function of parameters
well defined in the thermodynamic limit. These are the radius of the
sphere $R$ and the non-commutative parameter
$\Theta=2R^2/\sqrt{N^2-1}$ appearing in (\ref{eq:noncommutative}) and
in the list of possible limits of the fuzzy sphere of Table
\ref{tab:fuzzyspaces}, and $r$ and $\lambda$ appearing in the 
action (\ref{eq:accion2},\ref{eq:limitaccion}).

Using the scalings of (\ref{bcbar}) and $N\simeq 2R^2/\Theta$, we find
\begin{equation}
\bar{b}=\frac{r\Theta^{3/2}}{2\sqrt{2}\,R},\ \bar{c}=\frac{\lambda \Theta}{48\pi}.
\end{equation}
which can now be replaced in the algebraic fits for the coexistence
curves to get
\begin{eqnarray}
\mbox{Disorder/non-uniform:} & (\lambda\Theta)=122\left(-\frac{r\Theta
^{3/2}}{R}\right)-715 \\ 
\mbox{Disorder/uniform order:} &  (\lambda\Theta)=12.3\left(-\frac{r
\Theta^{3/2}}{R}\right) \\
\mbox{Uniform/non-uniform order:} &  (\lambda\Theta)=10.7\left(-
\frac{r\Theta^{3/2}}{R}\right)+10.6 \\ 
\mbox{Triple point:} & \left(-\frac{r_{_T}\Theta^{3/2}}{R}=6.5,\,
\lambda_{_T}\Theta=78.4\right). 
\end{eqnarray}
Since the phase boundary lines and the triple point all scale in the
same way,  it is not surprising to find that, out of the four
physical quantities available, only two are independent:
$r\Theta^{3/2}/R$ and $\lambda \Theta$. As a result there are not
enough physical parameters to fix the limiting procedure. For
instance,  if  the limiting space,  represented by $R$ and $\Theta$, is
fixed, one can still scale the field model parameters $r$ and
$\lambda$ freely.

\appendix

\section{Optimized algorithm for matrix models}\label{chap:optimized}
We now present an improved Monte-Carlo scheme we used to speed up our
simulations. 

The probability transition function (denoted by \ptf{} for short) of a
Monte-Carlo algorithm $W_{f,i}$ from an initial state $i$ with
probability $P_i$ to a final state $f$ with probability $P_f$, must
satisfy the detailed balance equation
\begin{equation}
P_i W_{f,i}=W_{i,f}P_f. \label{eq:metropol}
\end{equation}
The Metropolis \ptf{} 
\begin{equation}\label{eq:balance}
W_{fi}=\min \left[  1 , \frac{ P_{ f } }{ P_{ i }} \right] 
\end{equation}
is the best known example of one such, but another introduced in
\cite{boghosian} is given by
\begin{equation}\label{eq:bogho}
  W_{_{f,i}}^{\mathrm{B}} = w_{_{f,i}} \min \left[  1 , \frac{ P_{ f }
  }{ P_{ i }} \frac{w_{_{i,f}}}{ w_{_{f,i}}} \right] , 
\end{equation}
which is a generalization of the Metropolis \ptf{} when $w_{_{f,i}}\not=
w_{_{i,f}}$. In our case, we have selected the further generalization
\begin{equation}\label{eq:modif}
  w_{_{f,i}} = \min \left[ 1 , \frac{p_{f}}{p_{i}} \right],
\end{equation} 
which  is  equivalent  to  the  Metropolis \ptf{}  using  a  different
probability distribution $p$ yet to be defined.

The \emph{Boltzmann probability density distribution}\index{Boltzmann
  probability  density distribution} $P  \left( x  \right)$ to  find a
  configuration in the  volume $ \left( x, x +  \mathrm{d} x \right) $
  is defined by %
\begin{equation}\label{eq:boltz}
  P \left( x \right) = \frac{1}{Z} e^{-S \left[ x \right] },
\end{equation} 
where $S  \left[ x \right] $  is the action  or energy, and $Z$ is the
\emph{partition  function}\index{Partition function} which  contains
the whole  relevant information of the  system. In general,  it is not
possible  to obtain  an exact  expression for  the  partition function
analytically  or  numerically.  The  Monte--Carlo algorithm via  the
Metropolis \ptf{} (\ref{eq:metropol}) is so important because it does
not depend on $Z$. 

With $\Delta S = S \left[ x_{f} \right] - S \left[ x_{i} \right]$,
putting (\ref{eq:boltz}) in (\ref{eq:balance}) gives us the \ptf{} in
terms of the Boltzmann probability density distribution (denoted \pdd
for short), that is 
\begin{equation}\label{eq:metropoli}
  W_{{f,i}} = \mathrm{min} \left[ 1 , e^{- \Delta S} \right] .
\end{equation} 
In the same way as (\ref{eq:boltz}), we can associate a kind of energy
$s \left(y\right)$ to the probability distribution $p$ introduced in
(\ref{eq:modif}) 
\begin{equation}\label{eq:boltzsmall}
  p \left( y \right) = \frac{1}{z} e^{ -s \left[ y \right] }.
\end{equation}
Now,  the new \emph{Metropolis-Boghosian}\index{Metropolis-Boghosian}
\ptf{}    which   comes    from   the    equations
(\ref{eq:bogho}), (\ref{eq:modif}), (\ref{eq:boltz}) and (\ref{eq:boltzsmall}) is given by 
\begin{equation}\label{eq:boghofer}
  W_{_{f,i}}^{\mathrm{B}}  =  \mathrm{min} \left[  1  , e^{-\Delta  s}
  \right] \mathrm{min} \left[ 1 , e^{ -\Delta S + \Delta s} \right]
=\mathrm{min} \left[  1 , e^{-\Delta  s}  ,  e^{ -\Delta  S  },  e^{
  -\Delta  S +  \Delta  s} \right] , 
\end{equation} 
where $\Delta s=s \left[ x_{f} \right] - s \left[ x_{i} \right]$ is
the equivalent of the $\Delta S$ defined above. All the possible cases for
the \ptf{ } (\ref{eq:boghofer}) are presented in the
table~\ref{tab:bogho}.

\begin{table}[tbp]
\begin{center}
\begin{tabular}{|c|c|c|}
\hline
\null & $\Delta S \le 0$ &  $\Delta S > 0$ \\
\hline
$\Delta s \le 0  $ & $W_{_{f,i}}^{\mathrm{ B }} = \min  \left[ 1 , e^{
    -\Delta S +  \Delta s} \right] $ & $ W_{_{f,i}}^{  \mathrm{ B }} =
e^{ -\Delta S + \Delta s} $ \\
$\Delta s > 0 $ & $ W_{_{f,i}}^{ \mathrm{  B }} = e^{ - \Delta s } $ &
$ W_{_{f,i}}^{  \mathrm{ B }}  = \min  \left[ e^{ -  \Delta s }  , e^{
    -\Delta S   }  \right] $   \\  
\hline
\end{tabular}
\end{center}\caption{Possible   cases  in   the   evaluation  of   the
  Probability   Transition  Function   $W_{_{f,i}}^{\mathrm{   B  }}$.
  }\label{tab:bogho} 
\end{table}

We   can  ask  why  we might need  the   \ptf{}
(\ref{eq:boghofer})    when    we     have    a    simpler    function
(\ref{eq:metropoli}) already? When  the evaluation of $ \Delta  S $ is
quite  simple,  for instance  for the  Ising  model,  this methodology  is
counterproductive   because  more   exponential   functions  must   be
evaluated. On  the contrary, when  the evaluation of  $ \Delta S  $ is
computationally very expensive, as the  matrix models are, the \ptf{ }
(\ref{eq:boghofer}) avoids  the evaluation of  very improbable changes
in the configurations due to the implementation of the filter $ \Delta
s $, which reduces the processing time.

Numerically, we do not want to evaluate both $\Delta S$ and $\Delta
s$. If $s$ is a good approximation of the effective potential created
by $S$ but simpler to evaluate then, we can use the Metropolis
algorithm with the action $s$ to refuse or accept the new
configurations before the evaluation of $\Delta S$ (which is
complicated to evaluate and only will take machine time).

In this  section, we  propose a variant  calculation of  \ptf{}
(\ref{eq:boghofer}) shown in table~\ref{tab:boghofer}, where
  we avoid the  evaluation of $\Delta S$ when  the previous evaluation
  of Metropolis algorithm with  $\Delta s$ refuses the attempt to
  change the configuration.

\begin{table}[tbp]
\begin{center}
\begin{tabular}{|c|c|c|}
\hline
<\null & $\Delta S \le 0$ &  $\Delta S > 0$ \\
\hline
$\Delta s \le 0 $ & $W_{_{f,i}}^{\mathrm{ F }} = \min \left[ 1 , e^{ 
-\Delta S +  \Delta s} \right] $ &  $
W_{_{f,i}}^{ \mathrm{ F }} = e^{ -\Delta S +
    \Delta s}= \mathrm{min} \left[ 1 , e^{ -\Delta S +
    \Delta s} \right] $ \\ 
$\Delta s >  0 $ & $
W_{_{f,i}}^{ \mathrm{ F }} =  e^{ - \Delta s } $ &  $ W_{_{f,i}}^{
  \mathrm{ F }} = e^{ - \Delta s } $   \\  
\hline
\end{tabular}
\end{center}\caption{Proposition for a faster Probability Transition
  Function.}\label{tab:boghofer} 
\end{table}

 \subsection{Relative error}

The new \ptf, denoted by  $W_{_{f,i}}^{\mathrm{ F }}$, can not satisfy
the   detailed  balance  equation\index{Detailed balance equation}
(\ref{eq:balance}).   This  fact   introduces   deviations   in   the
probabilities, in  exchange for  a computational time  reduction, since
$\Delta s$ will be chosen simple to calculate. Anyway, we will make sure
to keep  under control  the error introduced  by this breaking  of the
detailed balance equation.

In accordance to this, the relative error between $W_{_{f,i}}^{ \mathrm{ B }}$ and
$W_{_{f,i}}^{ \mathrm{ F }}$ is defined as
\begin{equation}\label{eq:relativerror}
\mathrm{ err } = \left| 1 - \frac{ W_{_{f,i}}^{ \mathrm{ B }} }{
  W_{_{f,i}}^{ \mathrm{ F }} } \right|,
\end{equation}
and its values are shown in table~\ref{table:error}. As we can see in that
table, only the case $\Delta S > \Delta s > 0$ presents a relative error
different from zero. This error goes to zero when $\Delta S \gtrsim \Delta s$
and it goes to one when $ \Delta S \gg \Delta s$. Similarly, when $ \Delta S \gg
1$ we can almost take for granted the rejection of the new configuration by
Metropolis.  Thus, the introduction of the \ptf{ } $ W_{_{f,i}}^{ \mathrm{ F
}}$ is very convenient to estimate the \ptf{ } given by (\ref{eq:metropoli})
breaking the detailed balanced equation, where we only expect a
tolerably small deformation in the averages (specifically in regions with low
probability) with respect to the averages obtained from the \ptf's
(\ref{eq:metropoli}) and (\ref{eq:boghofer}).

\begin{table}[tbp]
\begin{center}
\begin{tabular}{|c|c|c|}
\hline
\null & $\Delta S \le 0$ &  $\Delta S > 0$ \\
\hline
$\Delta s \le 0 $ & $0$  &  $0$ \\
$\Delta s >  0 $ & $ 0$ & $\left\{\begin{tabular}{ll} 
0 & if  $\Delta S \le \Delta s$ \\
    $1 - e^{- \Delta S + \Delta s}$ &  if  $\Delta S > \Delta s $
  \end{tabular} \right. $\\  
\hline
\end{tabular}
\end{center}\caption{Relative error (\protect\ref{eq:relativerror})
  between the probability transition functions, $W_{_{f,i}}^{ \mathrm{
      B }}$ and $ W_{_{f,i}}^{ \mathrm{ F }}$.  }\label{table:error}
\end{table}

In our experience, we can obtain a better approximation when we
replace $\Delta s \to \left( \Delta S \right)^{\prime} $ only in the
case $\Delta S > \Delta s > 0$. The prime indicate the difference of
energy obtained by Metropolis one step before under the condition
$\Delta S > \Delta s > 0$. In that way, the algorithm has a kind of
``auto--regulation'' which reduces the deviation of the averages with
respect to the \ptf{ } which obeys the detailed balance equation.
With this auto--regulation we only update the energy reference level.

 \subsection{Application to the fuzzy scalar field
   model}\label{sec:matrixfuzzy} 
We can now adapt this scheme to the matrix model (\ref{eq:accion2}) we
are considering in this article.  A finite variation of the action,
from a configuration $\phi$ to a configuration $\phi +
\overline{\phi}$ is defined by $ \Delta S = S \left[ \phi +
\overline{\phi} \right] - S \left[ \phi \right]$, where
$\overline{\phi}$ must be a Hermitian matrix.  The evaluation of
$\left( \Delta S \right)_{\mu\nu}$ for the $\phi^{4}$ matrix model for
a single entry $ \left( \mu , \nu \right) $, involves the evaluation
of a cubic polynomial in the matrix. This is a highly non-local
function which causes the main slow--down in the code.

\begin{figure}[tbp]
  \begin{center}
\plotone{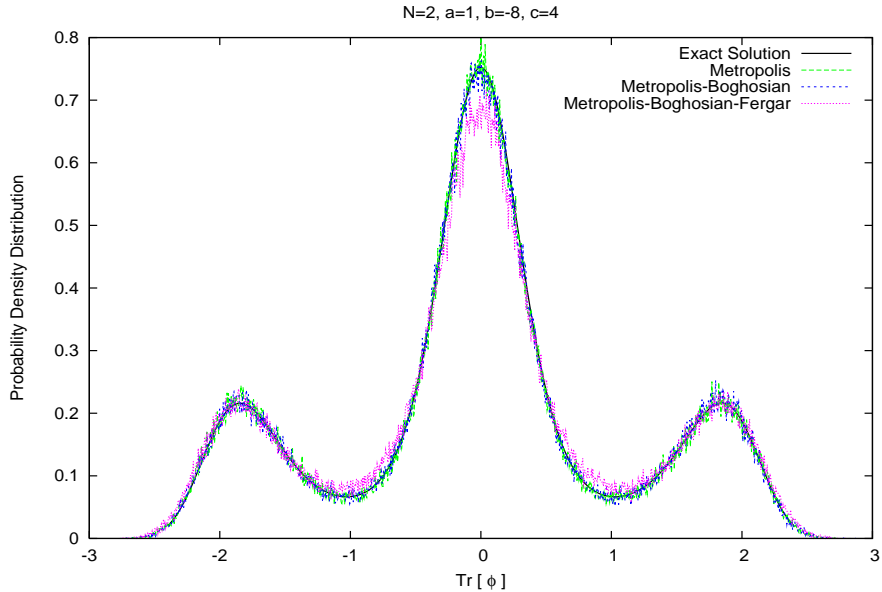}
\end{center}\caption{The  \pdd's  obtained  by different  methods  for
    $N=2$.}\label{fig:twoxtwodist} 
\end{figure}

Figure~\ref{fig:twoxtwodist} shows a comparison for the same set
of internal parameters (Monte Carlo time, thermalization time,
decorrelation time, etc.), at a collapsed point $(\bar{b},\bar{c})=(
-2^{3/2} , 1) $ from the phase diagram -Figure
\ref{fig:thephasediagram}- corresponding to the non-uniform ordered
phase, between the results obtained by direct integration (explained
in Section \ref{sec:lowest}), and the Monte Carlo simulations via
either of the three probability transition functions presented above:
Metropolis, Metropolis-Boghosian, and Metropolis-Boghosian-Fergar (the
one in Table \ref{tab:boghofer}). The small deviations (noise) of the
Monte--Carlo simulations with respect to the direct integration are a
normal effect of the Monte Carlo simulations and can be reduced by
increasing the number of samples produced by the code.

Going back to the choice of the filter $s$ introduced in
(\ref{eq:boltzsmall}), we want a function that incorporates the main
features of $S$ but is easier to evaluate.

We can notice that the action (\ref{eq:accion2}) for a
fixed entry $\left( \mu , \nu \right)$ of the matrix $\phi$ correspond
to a quartic polynomial in the entry $\phi_{\mu\nu}$, thus
\begin{equation}
  S_{_{\mu\nu}} = C \phi_{\mu\nu}^{4} + B \phi_{\mu\nu}^{2} + A, \nn 
\end{equation} 
where the coefficients $A$, $B$ and $C$ depend on the rest of the entries in
the matrix. In that sense, we propose
\begin{equation}\label{eq:miniaction}
  s_{_{\mu\nu}} = C^{\prime} \phi_{\mu\nu}^{4} + B^{\prime}
  \phi_{\mu\nu}^{2} + A^{ \prime} ,
\end{equation} 
where $A^{\prime}$, $B^{\prime}$ and $C^{\prime}$ are constant
coefficients. Thus, $s_{_{\mu\nu}}$ goes to $S_{_{\mu\nu}}$ when
$\left\{ A^{\prime} , B^{\prime} , C^{\prime} \right\} \to \left\{ A ,
B , C \right\}$. We will obtain a better concordance between both
$\Delta S$ and $\Delta s$ by choosing a set of
parameters $\left\{ A^{\prime} , B^{\prime}, C^{\prime} \right\}$
close to the non-primed parameters. As a first approximation, we took
  \begin{displaymath}
  \left\{ A^{\prime} , B^{\prime}, C^{\prime} \right\} =  \left\{
    \begin{array}{lcr}
\left\{ 0 , b , c \right\} & \mathrm{if } &\mu=\nu \\
\null \\
\left\{ 0 , 0 , c \right\} & \mathrm{if } &\mu\ne\nu \\
\end{array} \right.
\end{displaymath}
where  $b$ and  $c$ are  respectively  the bare  mass and  interaction
parameters  of the  model (\ref{eq:accion2}).   For simplicity  we have
fixed $A^{\prime}=0$ but in general,  we can use any other real number
and it will  not affect $\Delta s$. This set  of primed parameters for
$s$,  was chosen to  contain the  most basic  information of  the full
model $S$.

At the end of section~\ref{sec:modelonafuzzysphere}, we have shown
that the action~(\ref{eq:accion2}) has two symmetric minima with
respect to the trace. Those minima are located in $\mathrm{Tr} \left[
\phi \right] = \pm N \sqrt{- \frac{ b }{ 2 c } }$. Thus, \emph{we can
consider that every single diagonal entry in the matrix contributes to
the trace minima with $ \phi_{_{\mu \mu }} = \pm \sqrt{ - \frac{ b }{
2 c }} $, where ($ \mu = 1 , 2 , \dots , N $)}.  A simple function of
$ \phi_{_{\mu \mu}} $ with the same set of minima has been given in
the equation (\ref{eq:miniaction}) with the parameters $ \left(
A^{\prime } =0 , B^{ \prime } = b , C^{ \prime } = c \right) $.

For  non-diagonal  entries, we  have  observed  that their  probability
distribution  is  around zero.  Thus,  it  is  enough to  consider  the
function (\ref{eq:miniaction}) with  the parameters $ \left( A^{\prime
} =0 , B^{ \prime } = 0 , C^{ \prime } = c \right) $.  

\subsection{The algorithm}

\begin{figure}[tbp]
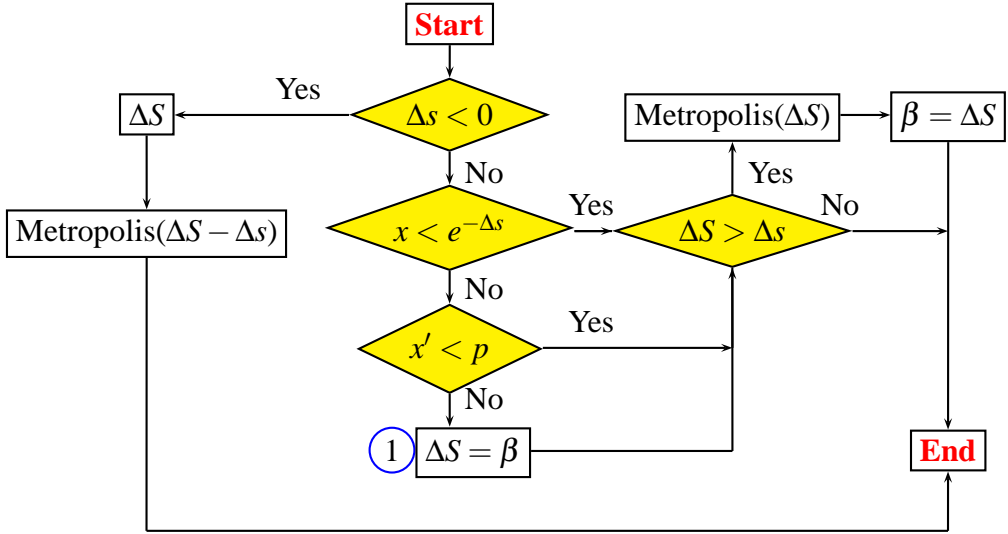

  \begin{center}
      \begin{psmatrix}[rowsep=0.4,colsep=0.5]
	{\null} & \psframebox{\textbf{\red Start}} &{\null}&{\null} \\
	\psframebox{$\Delta S$} &
	\psdiabox[fillstyle=solid,fillcolor=yellow]{$\Delta s <0$}
	&\psframebox{Metropolis($\Delta S$)}
	&\psframebox{$\beta = \Delta S$} \\
	\psframebox{Metropolis($\Delta S - \Delta s$)} &
	\psdiabox[fillstyle=solid,fillcolor=yellow]{$x < e^{-\Delta s}$}
	&\psdiabox[fillstyle=solid,fillcolor=yellow]{$\Delta S >
	\Delta s $}&{\null} \\
	{\null} &
	\psdiabox[fillstyle=solid,fillcolor=yellow]{$x^{\prime} < p $}
	&{\null}&{\null} \\
	{\null} & {\pscirclebox[linecolor=blue]{1}}\psframebox{$\Delta S = \beta$}
	&{\null}&\psframebox{\textbf{\red End}} \\
	{\null} & {\null}
	&{\null} &{\null}
%
%
%
	\ncline{->}{1,2}{2,2}
	\ncline{->}{2,2}{2,1}^{Yes}
	\ncline{->}{2,1}{3,1}
	\ncline{-}{3,1}{6,1}
	\ncline{-}{6,1}{6,4}
	\ncline{->}{6,4}{5,4}
	\ncline{->}{2,2}{3,2}>{No}
	\ncline{->}{3,2}{4,2}>{No}
	\ncline{->}{4,2}{5,2}>{No}
	\ncline{-}{5,2}{5,3}
	\ncline{->}{5,3}{3,3}
	\ncline{-}{3,2}{3,3}
	\ncline{->}{3,2}{3,3}^{Yes}
	\ncline{->}{3,3}{2,3}>{Yes}
	\ncline{->}{2,3}{2,4}
	\ncline{->}{2,4}{5,4}
	\ncline{->}{4,2}{4,3}^{Yes}
	\ncline{-}{4,3}{3,3}
	\ncline{->}{3,3}{3,4}^{No}
      \end{psmatrix}%
      %
      %
  \end{center}
\caption{Flow chart of the Optimized Monte--Carlo 
method.}
  \label{flow:MBF} 
\end{figure}

The algorithm is basically the same as the usual Metropolis algorithm
although with some adaptations to the current setting. In the
figure~\ref{flow:MBF} we show the flow chart for the implementation of
the new method that we have proposed.

%

%
%
%
%
\protect\index{Flow chart!optimazed Monte--Carlo method}

{\small
Internal variables in the flow chart \ref{flow:MBF}. 
\begin{multicols}{2}
\begin{itemize}
\item $x$  and $x^{\prime}$:  random numbers uniformly  distributed in
  the open interval $\left( 0, 1 \right) $. 
\item Metropolis($\Delta f$): indicates the Metropolis algorithm using
  the difference of energy $\Delta f$. 
\item $\beta$: represents the difference of the energy from a
  Monte--Carlo step which had been evaluated before.
\item $p$: ratio  of implementation of the new  method with respect to
  the standard one. 
\item {\pscirclebox[linecolor=blue]{1}}: part  of the subroutine where
  we avoid to evaluate $\Delta S$. 
\end{itemize}
\end{multicols}
}

In  the   flow  chart  presented   above,  we  have   emphasized  with
{\pscirclebox[linecolor=blue]{1}}, the step in the simulation where we
avoid the evaluation of $\Delta S$. This would seem to be insufficient
to reduce in a significant way  the processing time but it is not true
at  all.  The  number  of  times  that  the  algorithm  passes  trough
{\pscirclebox[linecolor=blue]{1}} divided by the total number of times
that the Modified Metropolis Algorithm (\mma{} for short) has been used,
will be an estimation of the efficiency of the new method with respect
to the Usual Metropolis Algorithm (\uma{} for short). 

Let us first define an efficiency parameter for our \mma{} to compare
it to the \uma. If $T_{_{MC}}$ is the Monte Carlo time to run the
simulation for $N \times N$ Hermitian matrices under the model
(\ref{eq:accion2}), and $\tau$ the number of times that the algorithm
passes trough the new feature {\pscirclebox[linecolor=blue]{1}}, then
\begin{equation}
  \mathrm{eff} = \frac{\tau}{N^{2}T_{_{MC}}}, \label{eq:eff}
\end{equation}
defines the efficiency of the modified algorithm. In particular, if
$t_{_{\mathrm{full}}}$ is the time for a run with a \uma{}, which is
without our routine {\pscirclebox[linecolor=blue]{1}} 
and $t_{_{\mathrm{gb}}}$ the time with it then, we have $t_{_{
\mathrm{ gb }}} \approx \left( 1 - \mathrm{ eff } \right) t_{_{
\mathrm{ full }}}$.

\emph{Remember  that  the figure  \ref{flow:MBF}  only represents  one
  attempt to switch one entry and,  if $ \beta $ has been updated then
  it must be saved for the next  one.} 

The modification of the Metropolis algorithm presented in this chapter
allows us to simulate matrix models with a decrease of the calculation
time with respect to the usual method. The explicit breaking of the
detailed balance equation by our proposition involves a systematic
error which we can keep under control at any time.

\subsection{The optimized Metropolis method}
As a second part, we present the results obtained with the optimized
Metropolis method. As explained in section \ref{sec:matrixfuzzy}, this
method was successfully tested in the lowest dimensional $N=2$ case.

As we saw in the figure~\ref{flow:MBF}, the Modified Metropolis
Algorithm (or \mma) had to evaluate three exponential functions
compared with the Usual Metropolis Algorithm (or \uma) where we only
have to evaluate one.  Thus, when the efficiency $\mathrm{eff}$
defined in (\ref{eq:eff}) is too small, it could be better to use the
\uma. This happens for instance, in the \emph{disordered
phase}\index{Disordered phase}: the difference in processing time
between \uma{} and \mma{} is not appreciable\footnote{They have
approximately the same velocity of processing because a large percent
of attempts will be in the range of fluctuations of $s$.}.  Even
worse, the processing time in \mma{} could be a little bigger in that
phase.

It is not the same for the other two phases where tunneling plays an
important role. There the efficiency $\mathrm{eff}$ goes to one and
the \mma{} is greatly more efficient\footnote{In these phases, the new
method is faster than the old one because a large percentage of
attempts could be out of the range of fluctuations of $s$, thus
avoiding the evaluation of $\Delta S$.}.

To keep under control the \emph{relative error}\index{Relative
error}\index{Error!relative} when $\mathrm{eff}$ is close to one, we
have to adjust the $p$ ratio. Thus, $p \approx 0$ means a
fast run, but could present a considerable relative error. At the
other extreme, when $p \approx 1$, the run will be slow but the
relative error will be very small.  We have to look for a balance
between accuracy and speed. We have set $p$ between $0.55$ and $0.70$
but it is also possible to set it dynamically.

\begin{figure}[tbp]
  \begin{center}
      \plotone{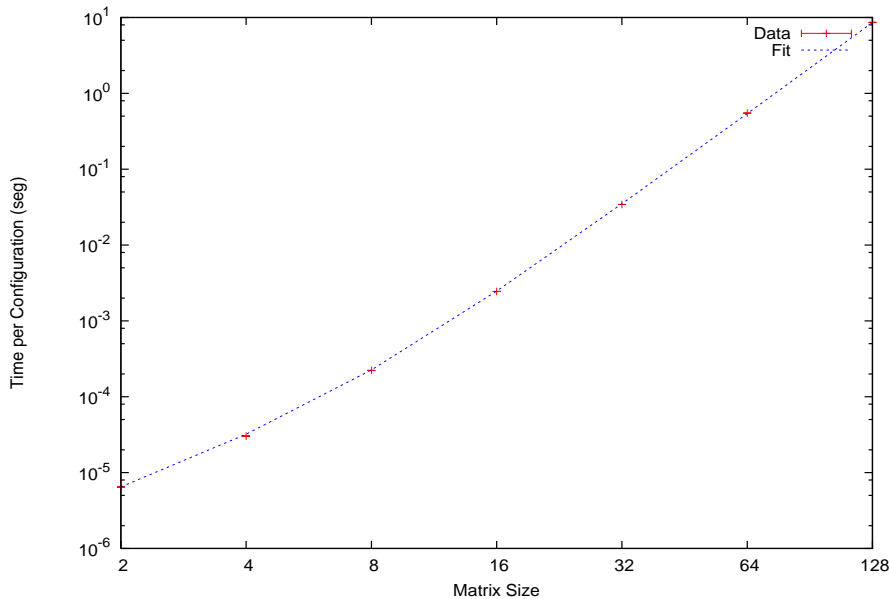} 
\end{center}  \caption{Processing  time   per  configuration  for  the
    \uma{}.}\label{fig:metropoltime} 
\end{figure}

As an example, in the figure~\ref{fig:metropoltime}, we show the
behavior of the processing time per configuration with respect to the
matrix size obtained by means of the usual Metropolis algorithm for
some given processor\footnote{In this example we have used a
\emph{Mobile Intel(R) Pentium(R) III CPU - M 800MHz, 369.10Mb RAM}.
On gcc-4.0.2 2005-10-01, Ubuntu, kernel 2.6.12-10-386. Kubuntu 5.10
Breezy Badger.} when $a=1$, $\bar{b}=-4$ and $\bar{c}=0.10$ which
corresponds to the \emph{uniform ordered phase} where we expect some
gain. And indeed, the best fit for this curve $\mathrm{time} = (1.49 \pm 0.02
\times 10^{ - 6 } ) N^{ 2 } + ( 3.18 \pm 0.02 \times 10^{-8})
N^{4}\,\mathrm{s}$ grows like $N^4$.

Starting from a random configuration, it can be thermalized or
decorrelated using the \mbf{} method described by Table
\ref{tab:boghofer}, then we can use the usual Metropolis algorithm to
evaluate the probability of transition between the old configuration
and this new sample obtained from \mbf. Doing this, we save processing time
and, at the same time, we do not introduce any systematic error
because the usual Metropolis algorithm will reject or accept the new
configuration which only contains the statistical error.

\end{document}